\documentclass{emulateapj}

\def\OH{$\log({\rm O/H})+12$}

\begin{document}
\slugcomment{Accepted to AJ}

\title{ MMT Extremely Metal Poor Galaxy Survey I.  An Efficient Technique to 
Identify Metal Poor Galaxies}

\author{Warren R.\ Brown}
\affil{Smithsonian Astrophysical Observatory}
\email{wbrown@cfa.harvard.edu}

\author{Lisa J. Kewley\altaffilmark{1}}
\affil{University of Hawaii}

\and

\author{Margaret J.\ Geller}
\affil{Smithsonian Astrophysical Observatory}

\altaffiltext{1}{Hubble Fellow}

\shorttitle{ MMT Extremely Low Metallicity Galaxy Survey I.}
\shortauthors{Brown, Kewley \& Geller}

\begin{abstract}

	We demonstrate a successful strategy for identifying extremely metal poor
galaxies.  Our preliminary survey of 24 candidates contains 10 metal poor galaxies
of which 4 have \OH$<7.65$, some of the lowest metallicity blue compact galaxies
known to date.  Interestingly, our sample of metal poor galaxies have systematically
lower metallicity for their luminosity than comparable samples of blue compact
galaxies, dIrrs, and normal star-forming galaxies.  Our metal poor galaxies share
very similar properties, however, with the host galaxies of nearby long-duration
gamma-ray bursts (GRBs), including similar metallicity, stellar ages, and star
formation rates.  We use H$\beta$ to measure the number of OB stars present in our
galaxies and estimate a core-collapse supernova rate of $\sim$10$^{-3}$ yr$^{-1}$.  
A larger sample of metal poor galaxies may provide new clues into the environment
where GRBs form and may provide a list of potential GRB hosts.

\end{abstract}

\keywords{
		galaxies: abundances ---
		galaxies: starburst ---
		gamma rays: bursts }

\clearpage

\section{INTRODUCTION}

	Metal poor galaxies are the key to understanding star formation and gas
enrichment in a nearly pristine interstellar medium, and may provide a template for
understanding the formation of the first stars.
	Extremely metal poor galaxies (XMPGs) are extremely rare:  fewer than 1\% of
dwarf galaxies are XMPGs, with a gas-phase oxygen abundance \OH $\leq 7.65$
\citep{kunth00, kniazev03}.  Known XMPGs are mostly gas-rich, blue compact galaxies
with spectra dominated by emission lines.  The first surveys to search for XMPGs
were objective prism surveys \citep{macalpine77, kunth81, terlevich91}.  Abundance
studies of these surveys revealed up to a dozen XMPGs \citep{kunth83, campbell86,
masegosa94}, but none so extreme as I Zw 18 \citep{searle72}.  I Zw 18, now along
with SBS 0335-052W and DDO 68, are the most metal poor local galaxies known, with
\OH\ ranging 7.12 - 7.17 \citep{izotov05, izotov07}.  \citet{papaderos06} found 2
new XMPGs in the 2dF survey.  Among the 1,000,000 spectra released by the Sloan
Digital Sky Survey (SDSS), there are 19 identified XMPGs \citep{kniazev03, izotov04,
izotov06b, izotov07}, emphasizing the rarity of such objects.

	\citet{kewley07} recently announced the serendipitous discovery of a new
XMPG, SDSS J080840.85+172856.48 (hereafter SDSS J0809+1729).  The object is a
stellar point source in the SDSS catalog.  It was observed as part of the
\citet{brown06} hypervelocity star survey on the basis of its stellar B-type
photometric colors.  Spectroscopy reveals that the object is not a star but rather a
compact blue galaxy at $cz=13232$ km s$^{-1}$.  Our re-analysis shows that the
galaxy has \OH$ =7.48 \pm 0.1$.  Interestingly, the observed electron density, star
formation rate, and total luminosity of this XMPG are remarkably similar to nearby
GRB host galaxies \citep{kewley07}.

	Nearby $z<0.2$ long duration GRBs are observed in metal poor galaxies
\citep{prochaska04, sollerman05, fruchter06, stanek06, wolf07, wiersema07,
margutti07}.  \citet{fruchter06} argue that the link between metal poor galaxies and
GRBs originates in the atmospheres of massive stars.  Massive metal poor stars lack
the opacity to support significant stellar winds, and thus can produce the anomalous
Type 1c supernovae associated with nearby GRBs.  \citet{berger07} argue that GRBs
are linked to young starburst populations, which at low redshift happen to be found
predominantly in low mass, metal-poor galaxies.  At large redshift $z>0.2$, the
metallicity of GRB hosts is much harder to determine \citep{prochaska06}.  At least
one GRB host galaxy is an extremely red and probably metal rich object
\citep{berger07}, and other GRB hosts $0.2<z<1$ appear consistent with normal
metallicity-luminosity relations \citep{wolf07, margutti07}.  GRBs clearly occur in
different types of host galaxies; they are not tied exclusively to the most metal
poor galaxies.  Yet finding and studying metal poor galaxies with properties similar
to nearby GRB host galaxies may yield critical clues about the environment where
nearby GRBs form.

	Inspired by the discovery of SDSS J0809+1729, we designed a survey to find
metal poor galaxies.  We use a technique similar to photometric redshifts to
identify metal poor galaxies in the SDSS galaxy catalog.  We test this technique and
target $g'\sim20$ galaxies with very blue colors, a region of parameter space not
well probed by previous surveys.  Our strategy uncovers 10 metal poor galaxies from
a sample of 24 candidates, 4 of which are new XMPGs.

	In \S 2 we present our technique to find new metal poor galaxies and discuss
the efficacy of our survey.  In \S 3 we describe the properties of the entire set of
metal poor galaxies, and compare the galaxies with samples of blue compact galaxies
and nearby GRB hosts.  In \S 4 we estimate the expected core-collapse supernova rate
in our galaxies.  We conclude in \S 5.

\begin{figure}		
 \includegraphics[width=3.25in]{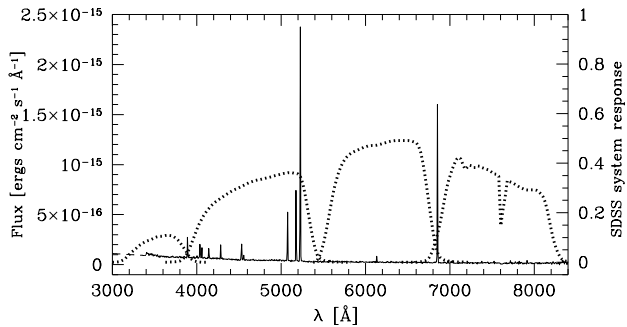}
 \caption{ \label{fig:filters} Observed spectrum of the XMPG SDSS J0809+1729
\citep{kewley07} plotted against the total system throughput of the SDSS
$u'~g'~r'~i'$ filter bandpasses (dotted lines).  The strong emission line
contribution to the galaxy's broadband magnitudes results in unusual colors that
vary with redshift. }
 \end{figure}

\section{DATA}

\subsection{Technique to Find Extremely Metal Poor Galaxies}

	Known XMPGs outside the Local Group are starburst galaxies characterized by
very low internal extinction, high ionization parameter, and large gas-phase
electron density \citep[e.g.][]{kniazev03, izotov05, izotov06b, papaderos06,
kewley07, izotov07}.  As a result, XMPGs have steep blue continua, large [O{\sc
iii}]/[O{\sc ii}] ratios, and strong hydrogen Balmer emission lines.  These
characteristics significantly affect the broadband colors of XMPGs.  For example,
Figure \ref{fig:filters} shows the observed spectrum of SDSS J0809+1729 plotted
against the total system throughput of the SDSS $u'g'r'i'$ filter
bandpasses\footnote{http://www.sdss.org/dr5/instruments/imager/index.html}.  The
contribution of emission lines to the broadband magnitudes of SDSS J0809+1729 is
approximately 3\%, 32\%, 6\%, and 12\% at $u'$, $g'$, $r'$, and $i'$ respectively.

	The strong emission line contribution to XMPGs' broadband magnitudes results
in unusual colors that change with redshift.  For example, at lower redshift, SDSS
J0809+1729 does not satisfy the B-star criteria in \citet{brown06b} because
H$\alpha$ drops into the $r'$ band and produces a much redder $(g'-r')$ color.  
Conversely, at higher redshift, H$\beta$ and \ion{O}{3} move from $g'$ into $r'$ and
also produce a much redder $(g'-r')$ color (see Figure \ref{fig:filters}).  We
quantify these effects using the RVSAO package \citep{kurtz98} to shift the XMPG
spectra to different redshifts. We predict the galaxy's colors at different
redshifts by adding the {\it relative} change in color to the observed photometry.  
In effect, we are calculating $k$-corrections for the XMPG.

\begin{figure}		
 \includegraphics[width=3.25in]{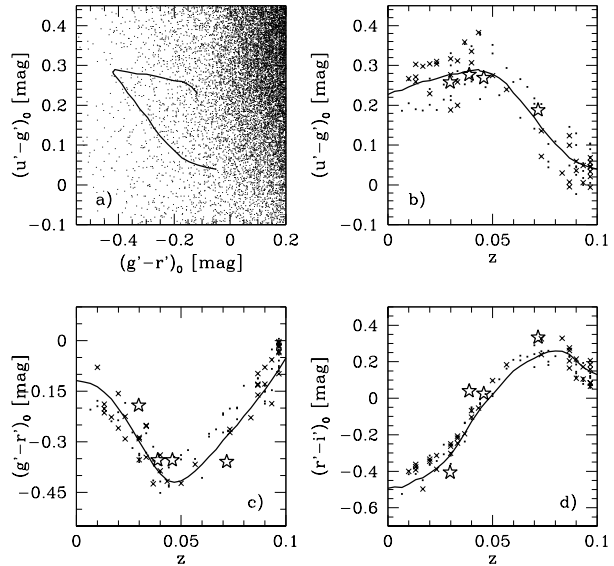}
 \caption{ \label{fig:tracks}
	Target selection for our MMT survey for XMPGs. a) Color-color plot showing
all SDSS DR4 galaxies (dots) and the color-redshift track for the XMPG J0809+1729
(solid line).  b) - d) Color-redshift plots showing SDSS objects (dots) that
simultaneously match all three color-redshift tracks (solid lines) within 0.1 mag.  
After removing saturated stars and nearby H{\sc ii} regions by visual inspection, we
are left with a sample of 38 XMPG candidates (x's).  Four new XMPGs discovered by
this survey are indicated by stars. }
 \end{figure}

	Figure \ref{fig:tracks} plots the resulting color-redshift tracks for SDSS
J0809+1729 (solid lines).  Over the range $0.0<z<0.10$, the XMPG's $(u'-g')_0$ and
$(g'-r')_0$ colors vary by $\sim0.3$ mag and the $(r'-i')_0$ color varies by $\sim$1
mag.  The subscript 0 indicates colors corrected for Galactic extinction following
\citet{schlegel98}.  We note that the Johnson/Cousins passbands are much less
sensitive than their SDSS equivalents because the strong H$\beta$/[\ion{O}{3}] and
H$\alpha$ emission lines remain in their respective $V$ and $R$ passbands for
$0.0<z<0.10$.

	Knowing how the broadband colors of an XMPG change with redshift, we can
search for new XMPGs at other redshifts in the SDSS photometric catalog.  
Unfortunately, searching for new XMPGs like SDSS J0809+1729 in the stellar catalog
is not feasible because of immense contamination from white dwarfs with similar
colors.  Instead, we search the SDSS galaxy catalog.

	We use the color-redshift track of our newly discovered XMPG, SDSS
J0809+1729, as the basis for the sample of XMPGs published here.  We begin by
selecting all galaxies in SDSS DR4 \citep{adelman06} with $g'<20.5$.  Figure
\ref{fig:tracks}a plots these galaxies with similar $(u'-g')_0$ and $(g'-r')_0$ to
SDSS J0809+1729.  Only by combining two or more colors can we meaningfully select
XMPG candidates.  We find $\sim10^3$ galaxies simultaneously within 0.1 mag of any
pair of color-redshift tracks, and a mere 107 galaxies (small squares, Figure
\ref{fig:tracks}b-d) simultaneously within 0.1 mag of all three color-redshift
tracks.  We visually inspect the objects and find that many are near saturated stars
or that they are HII regions in nearby galaxies.  After eliminating the unwanted
objects, we are left with a sample of 38 photometrically-selected XMPG candidates
(marked by x's, Figure \ref{fig:tracks}b-d).  Our survey is based on this sample of
38 XMPG candidates.

	We use a restrictive color selection as the first demonstration of our
technique.  If our XMPG selection strategy is successful, we can easily broaden the
search parameters to identify many more faint XMPG candidates.

\subsection{Observations}

	We obtained spectroscopy of the 24 candidates available on the nights of
2006 May 25-27 and 2006 June 19-20.  Table \ref{tab:other} lists the 24 candidates.  
Observations were obtained with the 6.5m MMT telescope and the Blue Channel
spectrograph.  We operated the spectrograph with the 300 line mm$^{-1}$ grating and
a 1$\arcsec$ slit.  These settings provide a wavelength coverage of 3400 \AA\ to
8600 \AA\ and a spectral resolution of 6.2 \AA.  Exposure times were typically 30
minutes.  We obtained comparison lamp exposures after every exposure.

	We reduce the spectra using standard IRAF\footnote{
        IRAF is distributed by the National Optical Astronomy Observatories, which
are operated by the Association of Universities for Research in Astronomy, Inc.,
under cooperative agreement with the National Science Foundation.}
	spectral reduction tasks and measure recession velocities from emission
lines using the package RVSAO \citep{kurtz98}.  We flux calibrate using
spectrophotometric standards \citep{massey88} and the standard Kitt Peak atmospheric
extinction correction.  For objects obtained in non-photometric conditions, we scale
the spectra by the flux ratio of the observed spectroscopic and SDSS broadband
magnitudes.  We estimate that absolute flux calibration is accurate to $\sim$25\%.

	We measure emission line fluxes using IRAF {\it splot} and {\it fitprof}
tasks, and find no significant offset between the two methods.  Table
\ref{tab:lineflux} presents the observed line strengths with their measurement
uncertainties.  The statistical uncertainties are formally a few percent, but the 
true error is dominated by uncertainties in the reddening correction, the stellar 
absorption correction, and the absolute flux calibration.

	For our analysis, we correct the observed emission line fluxes for reddening
using the Balmer decrement and the \citet{cardelli89} reddening curve.  We assumed
an $R_{V}=A_{V}/{\rm E}(\bv) = 3.1$ and an intrinsic H$\alpha$/H$\beta$ ratio of
2.85 \citep[the Balmer decrement for case B recombination at T$=10^4$K and $n_{e}
\sim 10^2 - 10^4 {\rm cm}^{-3}$;][]{osterbrock89}.

	We find no evidence for stellar absorption in the expected sense.  While one
would expect some underlying stellar absorption, we do not see Stark-broadened
absorption in the wings of the Balmer emission lines.  The de-reddened Balmer
emission line ratios exhibit a 7\% scatter around the \citet{osterbrock89} values
for case B recombination at T$=10^4$ K.  If we apply a constant 2 \AA\ equivalent
width correction, appropriate for the young 4-5 Myr stellar age of the galaxies (see
\S 3.4), the scatter of the Balmer line ratios around the \citet{osterbrock89}
values remains unchanged at 7\%.  We conclude that stellar absorption is smaller
than the uncertainty in the line ratio measurements.

	We calculate electron densities with the \ion{S}{2} $\lambda 6717$ /
\ion{S}{2} $\lambda 6731$ line ratio, when present, in conjunction with a 5 level
model atom using the Mappings photoionization code \citep{sutherland93}.  We derive
the gas-phase oxygen abundance following the procedure outlined in \citet{izotov06a}
within the framework of the classical two-zone HII-region model \citep{stasinska80}.  
This procedure utilizes the electron-temperature T$_{\rm e}$ calibrations of
\citet{aller84} and the atomic data compiled by \citet{stasinska05}.

	The gas-phase oxygen abundance depends on line ratios and thus is
independent of the uncertainties in our absolute flux calibration.  We propagate the
errors from the line ratio measurement, the extinction correction, and the stellar
absorption correction, and find that the relative errors in metallicities derived
using the same method are formally $\le 0.07$ dex.  However, the absolute error is
at least $\sim 0.1$ dex.  Thus systematics dominate the errors; our metallicities
are accurate at the $\pm0.1$ dex level.

\begin{figure}          
  \includegraphics[scale=0.89]{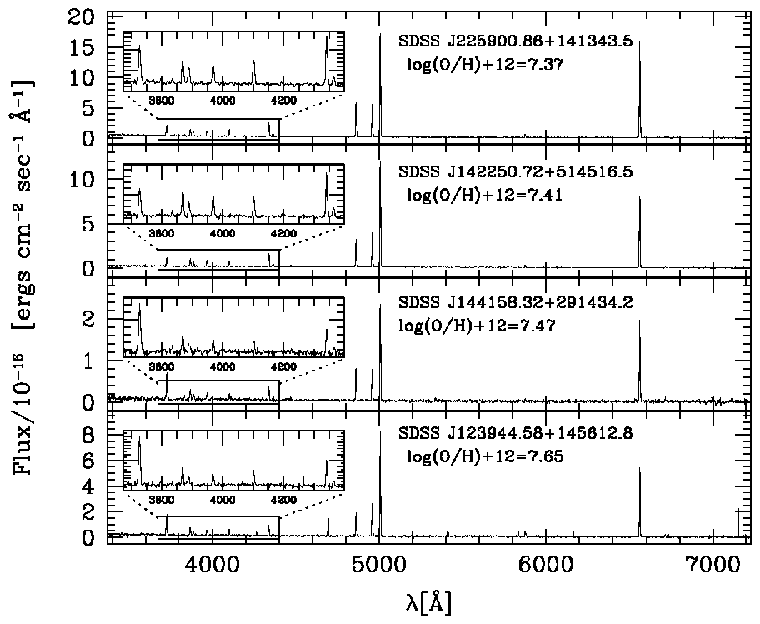}
  \includegraphics[scale=0.43]{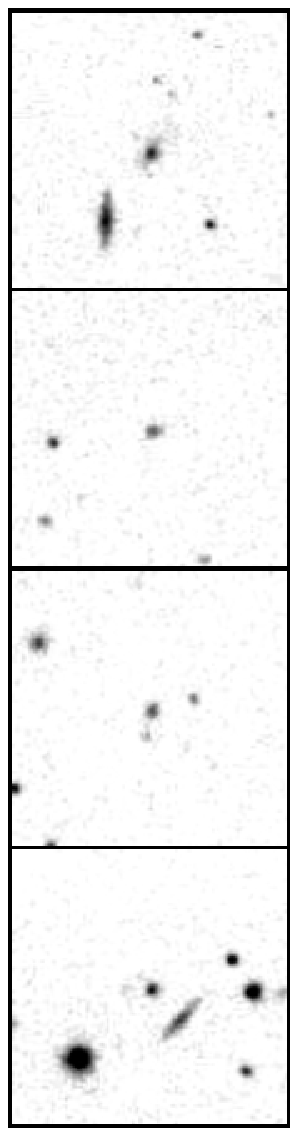}
 \caption{ \label{fig:spectra}
        MMT spectra of our 4 new XMPGs.  [\ion{O}{3}] $\lambda$4363 is well-detected
in all four objects (see insets).  The extremely weak [N{\sc ii}] and [S{\sc ii}]
lines visibly indicate the extremely low abundance of these objects.  The thumbnail
images, courtesy of SDSS, are 48\arcsec on a side and show that all 4 XMPGs appear
to be compact dwarfs.}
 \end{figure}

\subsection{Survey Efficiency}

	Our initial survey of 24 candidates contains 10 metal poor galaxies
with \OH $<8$ of which 4 are new XMPGs.  Thus our strategy is $\sim$20\% efficient
for selecting XMPGs.  Spectra and thumbnail images of the 4 new XMPGs are displayed
in Figure \ref{fig:spectra} (see also Table \ref{tab:gals}).  Three of the new XMPGs
(SDSS J142250.72+514516.5, SDSS J144158.32+291434.2 and SDSS J225900.86+141343.5)
have lower abundance than SDSS J0809+1729.  The remaining objects in our survey are
either A stars in the Milky Way, galaxies with modest emission lines, and a few odd
objects (one E+A, one possible BL Lac, one possible quasar) listed in Table
\ref{tab:other}.

\section{GALAXY PROPERTIES}

	We now open our discussion to include all the metal poor galaxies in our
survey.  Including SDSS J0809+1729, our survey contains 5 XMPGs and 6 metal poor
galaxies.  Table \ref{tab:gals} summarizes the spectroscopic measurements for the 11
galaxies.  Because the galaxies are compact objects, our spectra provide reasonable
global estimates of their properties.  We compare the properties of our galaxies
with similar samples of blue compact galaxies (BCGs) and with nearby GRB hosts.  
Throughout this paper we adopt a flat, $\Lambda$-dominated cosmology with $H_o=70$
and $\Omega_{m}=0.3$.

\begin{figure}		
 \includegraphics[width=3.0in]{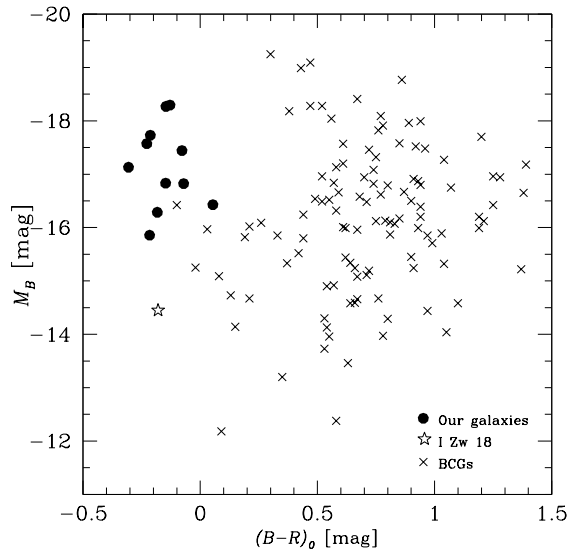}
 \caption{ \label{fig:colormag}
	Color-magnitude distribution of nearby BCGs \citep{paz03} and our 11 metal
poor galaxies, for which we estimate $(B-R)$ from SDSS photometry
\citep{fukugita96}.}
 \end{figure}

\subsection{Color and Redshift Distribution}

	Our 11 metal poor galaxies are systematically bluer and more luminous than
other known BCGs.  Figure \ref{fig:colormag} plots the \citet{paz03} sample of
nearby BCGs, for which $(B-R)_0$ and $M_B$ are all available.  We estimate $(B-R)$
for our galaxies from SDSS photometry \citep{fukugita96}, and shift the observed
magnitudes to the rest-frame using $k$-corrections we calculate for $B$ and $R$
passbands as described in \S 2.  It is clear that our metal poor galaxies are more
than 0.5 mag bluer in $(B-R)_0$ than most nearby BCGs.  I Zw 18 and UCM 1612+1308
\citep{rego98} are the two BCGs with colors comparable to our metal poor galaxies.  
However, our metal poor galaxies are systematically more luminous than the BCGs with
similar colors.

\begin{figure}		
 \includegraphics[width=3.0in]{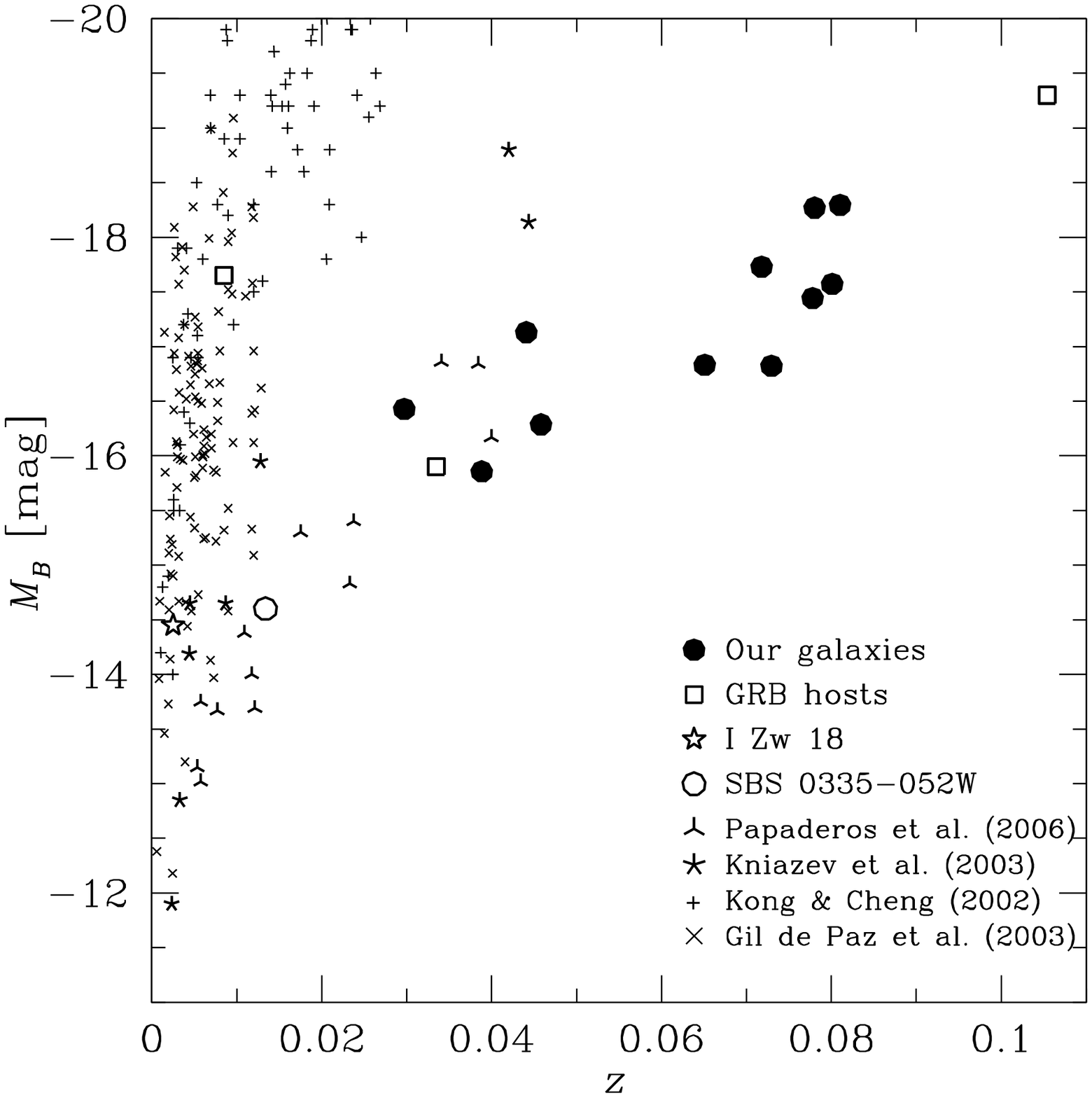}
 \caption{ \label{fig:zmag}
	Redshift and luminosity distribution of BCGs \citep{kong02,paz03},
metal poor galaxies from 2dF \citep{papaderos06} and SDSS \citep{kniazev03},
our 11 metal poor galaxies, and the 3 nearest GRB host galaxies
\citep{stanek06}.  $M_B$ is calculated for the 2dF galaxies assuming 
$(B-V)=0.5$ \citep{papaderos06}.  $M_B$ is estimated for the SDSS galaxies from 
SDSS photometry \citep{kniazev03}.}
 \end{figure}

	Our 11 metal poor galaxies are also at greater redshift than most known
BCGs.  Figure \ref{fig:zmag} plots the redshift distribution of BCGs
\citep{kong02,paz03}, metal poor galaxies from 2dF \citep{papaderos06} and SDSS
\citep{kniazev03}, our 11 metal poor galaxies, and the 3 nearest GRB host galaxies
\citep{stanek06}.  Note that we calculate $M_B$ for the 2dF galaxies assuming the
average $(B-V)=0.5$ for that sample \citep{papaderos06}.  We estimate $M_B$ for the
SDSS galaxies from $g'$ and $r'$ photometry in \citet{kniazev03}.  Figure
\ref{fig:zmag} shows that the vast majority of known BCGs have redshifts $z<0.02$.  
The 2dF and SDSS surveys access fainter magnitudes than earlier BCG samples and thus
contain a number of very low luminosity galaxies, plus a few higher luminosity metal
poor galaxies at $z\sim0.04$.  Our sample of metal poor galaxies, by construction,
spans the range $0.02<z<0.08$.  Searching a large volume of space enables the
discovery of rare objects like XMPGs.

\subsection{Luminosity-Metallicity Relation}

	Luminosity-metallicity relations are well determined for galaxies ranging
from large star-forming galaxies \citep[e.g.][]{tremonti04} to dwarf irregulars
\citep[e.g.][]{richer95}.  The physical basis for the luminosity-metallicity
relation is a mass-metallicity relation:  low mass galaxies are thought to sustain
less star formation and retain fewer metals than high mass galaxies.  We now compare
the luminosity and metallicity of our galaxies with samples of comparable metal-poor
dwarf galaxies.  Because there is some disagreement about the calibration of
different metallicity-estimate methods, we only consider galaxy samples with
metallicities derived with the T$_{\rm e}$ method.

\begin{figure}		
 \includegraphics[width=3.0in]{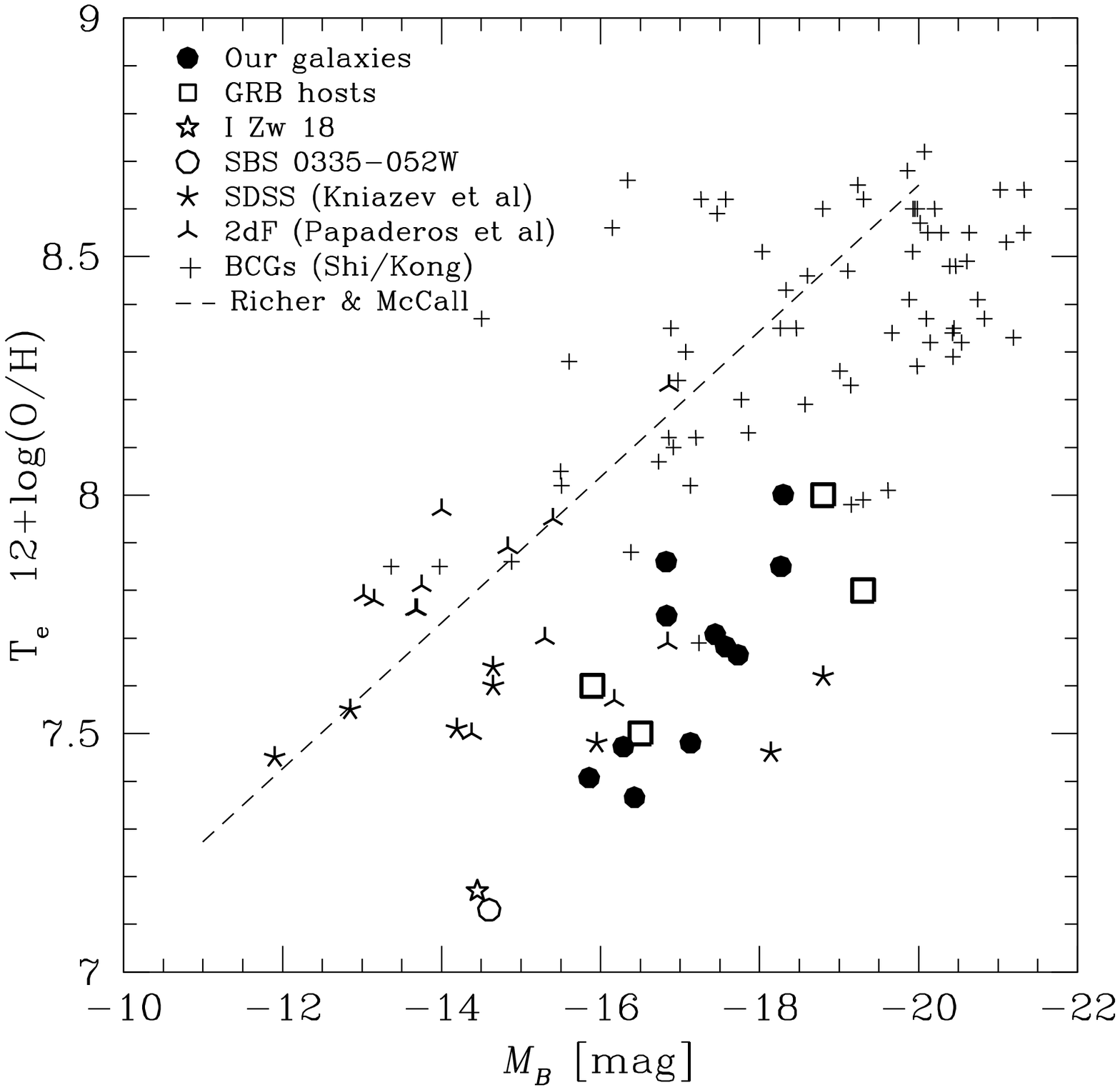}
 \caption{ \label{fig:lz}
	Luminosity-metallicity plot for BCGs \citep{kong02,shi05}, metal poor
galaxies from 2dF \citep{papaderos06} and SDSS \citep{kniazev03}, our metal poor
galaxies, and four nearby GRB host galaxies \citep{stanek06}.  Our sample of metal
poor galaxies fills in the region defined by the GRB hosts; all these galaxies have
lower metallicity by $\sim$0.5 dex than the \citet{richer95} luminosity-metallicity
relation ({\it dashed line}) for normal dIrrs.}
 \end{figure}

	Figure \ref{fig:lz} plots samples of BCGs \citep{kong02, shi05}, metal poor
galaxies from 2dF \citep{papaderos06} and SDSS \citep{kniazev03}, our metal poor
galaxies, and four nearby GRB host galaxies \citep{stanek06}.  We use metallicities
for the GRB hosts calculated using the T$_{\rm e}$ method as described in
\citet{kewley07}.  Figure \ref{fig:lz} shows that our sample of galaxies has either
1) lower metallicity by $\sim$0.5 dex than the \citet{richer95}
luminosity-metallicity relation for normal dIrrs, or 2) higher luminosity by 3 - 5
mag in $M_B$.

	Other samples of nearby, metal poor galaxies exhibit a large scatter around
the luminosity-metallicity relation \citep{kunth00}.  Yet our galaxies appear
unusual because of their large {\it systematic} offset from the
luminosity-metallicity relation.  We will use population synthesis models to address
the evolutionary paths of our metal poor galaxies in a future paper.

	Remarkably, our sample of metal poor galaxies fills the region of the
luminosity-metallicity diagram outlined by nearby GRB host galaxies.  The only other
galaxies with similar properties are the five metal poor 2dF and SDSS galaxies at
$z\sim0.04$:  2dF 169299, 2dF 115901, SDSSJ051902.64+000730.0,
SDSSJ104457.84+035313.2, and SDSSJ084030.00+470710.2.  The nearby BCG II Zw 70 also
appears to fall in the region defined by the GRB hosts.

\begin{figure}		
 \includegraphics[width=3.0in]{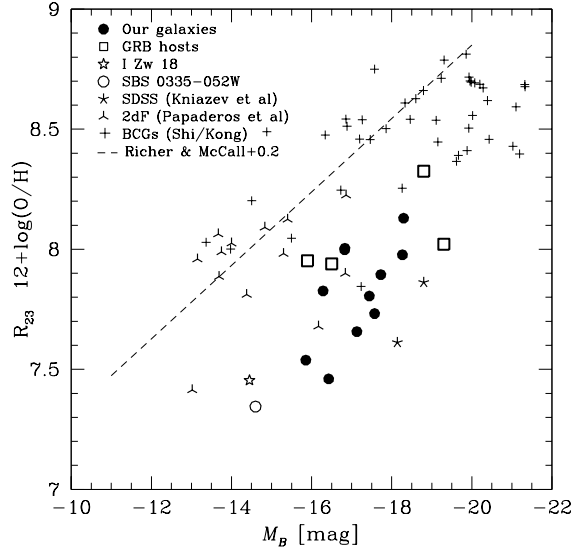}
 \caption{ \label{fig:lz2}
	Same as Figure \ref{fig:lz}, except we plot strong line metallicities
calculated using R$_{23}$ \citet{mcgaugh91}.  The strong line metallicities are
systematically +0.2 dex larger than the T$_{\rm e}$ metallicities, but the overall
distribution remains the same:  our sample of metal poor galaxies and the GRB hosts
are systematically offset from the \citet{richer95} luminosity-metallicity relation
({\it dashed line}), which we shift by +0.2 dex for consistency.}
 \end{figure}

	The relative distribution of galaxies in the luminosity-metallicity plot
does not change with different metallicity estimators.  To illustrate this point, we
calculate strong line metallicities using R$_{23}$ \citep{mcgaugh91} for the entire
set of galaxies.  The results are plotted in Figure \ref{fig:lz2}.  The R$_{23}$
metallicities have an average offset of +0.2 dex from the T$_{\rm e}$ metallicities,
and so we shift the \citet{richer95} luminosity-metallicity relation in Figure
\ref{fig:lz2} by a constant +0.2 dex for consistency.  Yet the relative distribution
of galaxies remains the same: our metal poor galaxies (and the GRB host galaxies)  
maintain a large, systematic offset from the luminosity-metallicity relation defined
by normal dIrrs and BCGs.

\subsection{Extinction}

	Our sample of 11 metal poor galaxies suffer from very little internal
extinction.  Table \ref{tab:gals} lists the Balmer H$\alpha$/H$\beta$ ratios, which
average $2.8\pm0.2$.  The intrinsic H$\alpha$/H$\beta$ ratio is 2.86 for case B
recombination at T$=10^4$ K and $n_e\sim10^2 - 10^4$ cm$^{-3}$ \citep{osterbrock89}.  
Low extinction is expected in low metallicity galaxies, and is also observed in
nearby GRB hosts \citep[e.g.][]{kewley07}.

\subsection{Star Formation Age and Rates}

	XMPGs remain a puzzle because they may be pristine galaxies undergoing their
first burst of star formation or they may contain older stellar population from
previous episodes of star formation.  {\it Hubble Space Telescope} images resolve
old stellar populations in the nearest metal poor galaxies.  I Zw 18, for example,
has stars with ages ranging from $\sim 500$~Myr \citep{izotov04b} to $\sim 1$~Gyr
\citep{aloisi99}.  Although we cannot estimate the age of old stellar populations
(if any) in our metal poor galaxies, stellar population synthesis models provide an
estimate of the age of the young stellar population.

	We estimate the stellar age of our metal poor galaxies from the H$\beta$
equivalent width following \citet{schaerer98}.  Under the assumption of a Salpeter
initial mass function and an instantaneous burst of star formation, the stellar ages
of our metal poor galaxies are in the range 4 - 5 Myr (see Table \ref{tab:gals}).

	Star formation rates (SFRs) are more difficult to estimate because
traditional calibrations are not applicable to extremely metal poor objects.  We use
the calibration derived by \citet{kewley07} based on the stellar population
synthesis models of \citet{bicker05}.  Our metal poor galaxies then have SFRs in the
range 0.1 - 0.5 M$_{\sun}$ yr$^{-1}$ (Table \ref{tab:gals}).  Both the stellar ages
and SFRs of our metal poor galaxies are remarkably similar to those found for nearby
GRB hosts \citep{kewley07}.

\section{SUPERNOVA RATES AND THE GRB CONNECTION}

	We next ask how frequently GRBs might occur in our sample of metal poor
galaxies, if GRBs are indeed associated with core-collapse supernova in metal poor
galaxies.  We start by estimating the number of massive O and B stars in our sample
of 11 galaxies, and then estimate the rate of core-collapse supernovae.

	Under the assumption of an ionization-bounded nebula, the H$\beta$ line
luminosity provides an estimate of the ionizing flux present in the galaxies.  The
metal poor galaxies suffer from very little extinction, as measured by their
H$\alpha$/H$\beta$ ratios; thus we make no correction for internal extinction.
	We follow \citet{schaerer98} and convert the observed H$\beta$ ionizing flux
to an equivalent number of O stars.  We select the ``equivalent O7V to O star''
ratio $\eta_o$ based on the stellar age and metallicity of the galaxy; $\eta_o$
values range from 0.25 - 0.5.  The minimum O star mass is $\sim$13.3 M$_\sun$ at our
metallicities \citep{vacca94}.  However, core-collapse supernovae will result from B
stars with masses as low as $\sim$8 M$_\sun$.  A Salpeter initial mass function has
the same number of O stars with 13 - 120 M$_\sun$ as B stars with 8 - 13 M$_\sun$.  
Thus, the total number of core-collapse supernova progenitors is roughly twice the
\citet{schaerer98} number of equivalent O stars.  We list our estimate of the total
number of core-collapse supernova progenitors in the column n(OB) (Table
\ref{tab:gals}).  On average, there are $\sim$25,000 potential core-collapse
supernova progenitors per galaxy in our sample.  This number may appear relatively
small, but the metal poor galaxies are dwarfs, not massive galaxies.

	We estimate core-collapse supernova rates in our metal poor galaxies by
assuming supernovae occur uniformly over the lifetime of the longest-lived
progenitor.  The \citet{schaller92} stellar evolution track for a 9 M$_\sun$ star
with $Z=0.001$ has a lifetime of 30 Myr.  Thus the average rate of core-collapse
supernovae in our metal poor galaxies is $\sim$10$^{-3}$ yr$^{-1}$, with an
uncertainty of a factor of a few.  In other words, a couple thousand such metal poor
galaxies must be monitored to witness one core-collapse supernova per year.  A
typical spiral galaxy, in comparison, has a typical SFR of $\sim$10 M$_\sun$
yr$^{-1}$ \citep[e.g.][]{brinchmann04} and a core-collapse supernova rate $\sim$100
times larger than the metal poor galaxies, assuming a constant SFR and Salpeter
initial mass function.

	The ratio of GRBs to core-collapse supernovae events depends on how narrowly
beamed GRBs are.  If 1 out of 100 core-collapse supernovae appear as GRBs, then a
sample of a $\sim$10$^5$ metal poor galaxies must be monitored to witness one GRB
per year.  There is currently little constraint on the space density of metal poor
galaxies, in part because they are such low-luminosity systems.  Surveys reaching
faint magnitude limits may uncover large numbers of new metal poor galaxies.  
However, it may be difficult to detect a supernova coincident on the high
surface-brightness core of these galaxies.  Figure \ref{fig:spectra} shows that the
four XMPGs, for example, are compact systems.  Although metal poor galaxies are
potential GRB hosts, a survey to find GRBs by monitoring a sample metal poor
galaxies appears impractical currently.  Future deep imaging surveys, such as
Pan-STARRS and the Large Synoptic Survey Telescope (LSST), may be able to detect
supernovae in these low-luminosity galaxies.

\section{CONCLUSIONS}

	We have designed a successful strategy to find new metal poor galaxies.  We
calculate the expected colors of metal poor galaxies at different redshifts based on
the spectrum of the newly discovered XMPG, SDSS J0809+1729 \citep{kewley07}.  We
observed an initial sample of 24 candidates with the MMT telescope and find 4 new
XMPGs with \OH $\leq 7.65$, a $\sim$20\% selection efficiency.

	Our full set of 11 metal poor galaxies are systematically bluer and more
luminous than comparable samples of BCGs.  Our galaxies are also systematically more
metal poor by 0.5 dex (or more luminous by 3 - 5 mag) than samples of BCGs, dIrrs,
and other metal-poor dwarf galaxies.  Remarkably, our galaxies share the same region
of the luminosity-metallicity diagram with nearby GRB hosts.  The similarity in
extinction, stellar age, and star formation rates suggests that our metal poor
galaxies are potential hosts for GRBs.

	We estimate an average core-collapse supernova rate $\sim$$10^{-3}$
yr$^{-1}$ in our metal poor galaxies.  This estimate comes from an estimate of the 
number of O and B stars in the galaxies.  If GRBs are indeed linked to core-collapse 
supernova in metal poor galaxies, future surveys such as Pan-STARRS or LSST may be 
able to find GRBs by monitoring a large sample of metal poor galaxies.

	The success of our XMPG selection strategy allows us to expand our survey.  
For example, using the new XMPGs (Figure \ref{fig:spectra}) as additional templates,
we identify a total of 335 XMPG candidates in SDSS Data Release 5.  Spectroscopic
observations of these XMPG candidates are underway.

\acknowledgements

	W.~R.~Brown was supported in part by a Clay Fellowship during this work.  
L.~J.~Kewley was supported by a Hubble Fellowship.  We thank S.\ Kenyon and the
anonymous referee for comments that greatly improved this paper.  This research has
made use of NASA's Astrophysics Data System Bibliographic Services.  This project
makes use of data products from the Sloan Digital Sky Survey, which is managed by
the Astrophysical Research Consortium for the Participating Institutions.  We thank
the Smithsonian Institution for partial support of this research.

{\it Facilities:} MMT (Blue Channel Spectrograph)




\begin{deluxetable}{lcrl}           
\tablecolumns{4} 	\tablewidth{0pt}
\tablecaption{XMPG CANDIDATES\label{tab:other}}
\tablehead{
  \colhead{ID} & \colhead{$g'$} & \colhead{$cz$} & \colhead{Comment} \\
  \colhead{} & \colhead{(mag)} & \colhead{(km s$^{-1})$} & \colhead{}
}
	\startdata
SDSS J100539.35+315441.6 & 18.70 &   ... & quasar?           \\
SDSS J120955.68+142155.7 & 20.17 & 23340 & metal poor galaxy \\
SDSS J123944.58+145612.8 & 19.77 & 21534 & metal poor galaxy \\
SDSS J124638.82+350115.1 & 20.36 & 19522 & metal poor galaxy \\
SDSS J124709.24+325118.6 & 19.10 & 27450 & galaxy            \\
SDSS J133424.53+592057.0 & 20.55 & 21890 & metal poor galaxy \\
SDSS J135641.64+654748.8 & 19.72 & 10742 & galaxy            \\
SDSS J140439.28+542136.9 & 19.98 &   318 & H{\sc ii} region  \\
SDSS J141333.54+463234.0 & 19.43 &    90 & A-star            \\
SDSS J142250.72+514516.5 & 20.22 & 11654 & metal poor galaxy \\
SDSS J143345.99-025602.2 & 19.69 &    65 & A-star            \\
SDSS J144158.32+291434.2 & 20.13 & 13741 & metal poor galaxy \\
SDSS J145621.69+503523.0 & 20.49 &   ... & BL Lac?           \\
SDSS J150316.52+111056.9 & 19.41 & 23405 & metal poor galaxy \\
SDSS J150535.89+314639.4 & 20.55 & 15817 & galaxy            \\
SDSS J151221.08+054911.2 & 20.15 & 24025 & metal poor galaxy \\
SDSS J152802.62+240425.6 & 18.65 &    -6 & A-star            \\
SDSS J154742.23-005554.2 & 20.57 & 16594 & E+A galaxy        \\
SDSS J160238.71+444923.8 & 19.65 & 12350 & galaxy            \\
SDSS J165835.08+192415.3 & 18.93 &   -23 & A-star            \\
SDSS J172955.61+534338.8 & 19.55 & 24308 & metal poor galaxy \\
SDSS J211613.97-000851.3 & 20.39 & 11916 & galaxy            \\
SDSS J221912.56+140602.8 & 20.41 &    36 & A-star            \\
SDSS J225900.86+141343.5 & 19.09 &  8918 & metal poor galaxy \\
	\enddata
 \end{deluxetable}

\begin{deluxetable}{lcccccccccc}
\tablecolumns{11}	\tablewidth{0pt}
\tabletypesize{\scriptsize}
\tablecaption{OBSERVED LINE INTENSITIES\label{tab:lineflux}}
\tablehead{
 \colhead{F($\lambda_0$ Ion)\tablenotemark{a}} &
 \colhead{J120955.67} & \colhead{J123944.58} & \colhead{J124638.82} &
 \colhead{J133424.53} & \colhead{J142250.72} & \colhead{J144158.32} &
 \colhead{J150316.52} & \colhead{J151221.08} & \colhead{J172955.61} & 
 \colhead{J225900.86} \\
 \colhead{} &
 \colhead{+142155.9} & \colhead{+145612.8} & \colhead{+350115.1} &
 \colhead{+592057.0} & \colhead{+514516.5} & \colhead{+291434.2} &
 \colhead{+111056.9} & \colhead{+054911.2} & \colhead{+534338.8} &
 \colhead{+141343.5}
}
	\startdata
3727 [O{\sc ii}]  &  8.20 $\pm$ 0.22 &  11.8 $\pm$  0.2 &  16.4 $\pm$  0.4 &  22.5 $\pm$  0.3 &  7.80 $\pm$ 0.14 &  5.70 $\pm$ 0.13 &  43.8 $\pm$  0.5 &  9.00 $\pm$ 0.12 &  29.9 $\pm$  0.5 &  15.1 $\pm$  0.2 \\
3798 H10          &      \nodata     &  0.84 $\pm$ 0.10 &  0.49 $\pm$ 0.06 &  0.62 $\pm$ 0.07 &  0.35 $\pm$ 0.07 &  0.27 $\pm$ 0.05 &  0.72 $\pm$ 0.09 &  0.36 $\pm$ 0.10 &  1.25 $\pm$ 0.11 &  1.20 $\pm$ 0.16 \\
3835 H9           &      \nodata     &  0.29 $\pm$ 0.08 &  0.34 $\pm$ 0.06 &  0.55 $\pm$ 0.07 &  0.91 $\pm$ 0.10 &  0.45 $\pm$ 0.07 &  1.35 $\pm$ 0.12 &  0.35 $\pm$ 0.06 &  1.38 $\pm$ 0.11 &  1.31 $\pm$ 0.15 \\
3868 [Ne{\sc iii}]&  1.75 $\pm$ 0.20 &  3.38 $\pm$ 0.14 &  4.09 $\pm$ 0.13 &  5.62 $\pm$ 0.11 &  6.11 $\pm$ 0.27 &  1.47 $\pm$ 0.12 &  14.5 $\pm$  0.3 &  2.98 $\pm$ 0.23 &  13.5 $\pm$  0.4 &  7.07 $\pm$ 0.12 \\
3889 He {\sc i} + H8    &  0.66 $\pm$ 0.08 &  2.12 $\pm$ 0.17 &  1.86 $\pm$ 0.21 &  2.22 $\pm$ 0.11 &  3.14 $\pm$ 0.11 &  0.84 $\pm$ 0.10 &  5.43 $\pm$ 0.11 &  1.44 $\pm$ 0.11 &  5.09 $\pm$ 0.29 &  5.52 $\pm$ 0.11 \\
3968 [Ne{\sc iii}] + H7 &  1.07 $\pm$ 0.11 &  2.50 $\pm$ 0.11 &  2.39 $\pm$ 0.12 &  3.84 $\pm$ 0.14 &  4.78 $\pm$ 0.28 &  0.89 $\pm$ 0.09 &  8.67 $\pm$ 0.15 &  1.90 $\pm$ 0.11 &  8.21 $\pm$ 0.11 &  6.80 $\pm$ 0.12 \\
4101 H$\delta$    &  1.40 $\pm$ 0.11 &  2.50 $\pm$ 0.12 &  2.40 $\pm$ 0.23 &  3.60 $\pm$ 0.18 &  3.80 $\pm$ 0.11 &  0.90 $\pm$ 0.11 &  8.50 $\pm$ 0.19 &  2.30 $\pm$ 0.13 &  6.90 $\pm$ 0.14 &  7.90 $\pm$ 0.11 \\
4340 H$\gamma$    &  2.70 $\pm$ 0.10 &  4.30 $\pm$ 0.11 &  4.40 $\pm$ 0.10 &  6.10 $\pm$ 0.10 &  8.30 $\pm$ 0.11 &  1.90 $\pm$ 0.11 &  15.9 $\pm$  0.3 &  4.00 $\pm$ 0.11 &  12.4 $\pm$  0.2 &  15.0 $\pm$  1.0 \\
4363 [O{\sc iii}] &  0.30 $\pm$ 0.04 &  1.10 $\pm$ 0.13 &  1.00 $\pm$ 0.11 &  1.10 $\pm$ 0.10 &  2.00 $\pm$ 0.20 &  0.40 $\pm$ 0.07 &  3.50 $\pm$ 0.14 &  0.70 $\pm$ 0.09 &  2.90 $\pm$ 0.19 &  2.90 $\pm$ 0.11 \\
4471 He {\sc i}   &      \nodata     &      \nodata     &  0.36 $\pm$ 0.06 &  0.32 $\pm$ 0.04 &  0.73 $\pm$ 0.09 &  0.29 $\pm$ 0.04 &  1.31 $\pm$ 0.11 &      \nodata     &      \nodata     &  0.66 $\pm$ 0.08 \\
4686 He {\sc ii}  &      \nodata     &      \nodata     &  0.19 $\pm$ 0.05 &  0.23 $\pm$ 0.04 &  0.73 $\pm$ 0.09 &      \nodata     &      \nodata     &  0.37 $\pm$ 0.05 &  0.41 $\pm$ 0.09 &  0.68 $\pm$ 0.08 \\
4861 H$\beta$     &  5.50 $\pm$ 0.11 &  10.4 $\pm$  0.1 &  10.3 $\pm$  0.1 &  14.1 $\pm$  0.1 &  18.3 $\pm$  0.2 &  4.00 $\pm$ 0.10 &  37.9 $\pm$  0.4 &  9.80 $\pm$ 0.15 &  29.2 $\pm$  0.4 &  33.4 $\pm$  0.3 \\
4959 [O{\sc iii}] &  5.50 $\pm$ 0.24 &  15.1 $\pm$  0.4 &  14.7 $\pm$  0.1 &  20.8 $\pm$  0.2 &  21.8 $\pm$  0.2 &  4.50 $\pm$ 0.21 &  62.8 $\pm$  0.6 &  11.9 $\pm$  0.2 &  62.5 $\pm$  0.7 &  33.5 $\pm$  0.3 \\
5007 [O{\sc iii}] &  16.3 $\pm$  0.2 &  45.0 $\pm$  0.5 &  44.6 $\pm$  0.5 &  61.8 $\pm$  0.7 &  65.5 $\pm$  0.7 &  13.6 $\pm$  0.2 & 190.0 $\pm$  1.9 &  35.2 $\pm$  0.4 & 188.0 $\pm$  2.2 & 100.0 $\pm$  1.0 \\
5876 He {\sc i}   &  0.46 $\pm$ 0.06 &  2.60 $\pm$ 0.11 &  1.18 $\pm$ 0.11 &  1.73 $\pm$ 0.12 &  2.08 $\pm$ 0.32 &  0.37 $\pm$ 0.05 &  3.73 $\pm$ 0.12 &  1.03 $\pm$ 0.11 &  3.76 $\pm$ 0.13 &  2.80 $\pm$ 0.13 \\
6563 H$\alpha$    &  14.8 $\pm$  0.2 &  30.7 $\pm$  0.4 &  31.5 $\pm$  0.3 &  40.3 $\pm$  0.7 &  46.2 $\pm$  0.5 &  11.3 $\pm$  0.1 & 102.0 $\pm$  1.2 &  25.8 $\pm$  0.3 &  94.2 $\pm$  1.0 &  91.9 $\pm$  0.9 \\
6584 [N{\sc ii}]  &      \nodata     &      \nodata     &  0.70 $\pm$ 0.08 &  1.30 $\pm$ 0.21 &      \nodata     &  0.20 $\pm$ 0.03 &  1.90 $\pm$ 0.11 &  0.40 $\pm$ 0.05 &  1.40 $\pm$ 0.11 &  0.40 $\pm$ 0.09 \\
6678 He {\sc i}   &      \nodata     &      \nodata     &      \nodata     &  0.39 $\pm$ 0.06 &      \nodata     &      \nodata     &  0.94 $\pm$ 0.25 &      \nodata     &  0.83 $\pm$ 0.09 &  0.42 $\pm$ 0.07 \\
6717 [S{\sc ii}]  &  1.40 $\pm$ 0.26 &  1.00 $\pm$ 0.16 &  1.40 $\pm$ 0.11 &  2.20 $\pm$ 0.11 &  0.70 $\pm$ 0.08 &  0.50 $\pm$ 0.09 &  3.70 $\pm$ 0.12 &  0.90 $\pm$ 0.10 &  3.20 $\pm$ 0.14 &  1.40 $\pm$ 0.14 \\
6731 [S{\sc ii}]  &  0.50 $\pm$ 0.08 &  1.00 $\pm$ 0.15 &  0.90 $\pm$ 0.13 &  1.30 $\pm$ 0.12 &  0.50 $\pm$ 0.09 &  0.30 $\pm$ 0.06 &  3.20 $\pm$ 0.11 &      \nodata     &  2.20 $\pm$ 0.12 &  1.00 $\pm$ 0.11 \\
EW(H$\beta$)      &  54.4 $\pm$  3.0 &  91.2 $\pm$  5.0 &  70.7 $\pm$  3.9 &  72.5 $\pm$  4.0 &  95.9 $\pm$  5.3 &  76.7 $\pm$  4.2 &    80 $\pm$  4.4 &  61.6 $\pm$  3.4 &   158 $\pm$  8.7 &   134 $\pm$  7.4 \\
	\enddata
 \tablenotetext{a}{Flux in units of $\times10^{-16}$ ergs s$^{-1}$ cm$^{-2}$.  
Uncertainties are statistical errors only, and do not include the errors due to 
reddening, stellar absorption, and absolute flux calibration.}
 \end{deluxetable}

\begin{deluxetable}{lccccccccrc}           
\tablecolumns{11} 	\tablewidth{0pt}
\tabletypesize{\scriptsize}
\tablecaption{OUR METAL POOR GALAXIES\label{tab:gals}}
\tablehead{
  \colhead{ID} & \colhead{$cz$} & \colhead{$M_B$} & \colhead{T$_e$} & \colhead{\OH}
  & \colhead{Age} & \colhead{H$\alpha$/H$\beta$}
  & \colhead{L(H$\beta$)} & \colhead{SFR (T$_e$)} & \colhead{N(OB)}
  & \colhead{SN rate} \\
  \colhead{} & \colhead{(km s$^{-1})$} & \colhead{(mag)} & \colhead{(K)} & \colhead{}
  & \colhead{(Myr)} & \colhead{}
  & \colhead{(erg s$^{-1}$)} & \colhead{(M$_{\sun}$ yr$^{-1}$)}
  & \colhead{} & \colhead{(10$^{-4}$ yr$^{-1}$)}
}
	\startdata
SDSS J225900.86+141343.51 &  8918 & -16.4 & 18600 & 7.37 & 4.0 & 2.75 & 1.80$\times10^{40}$ & 0.21 &  13900 &  4 \\
SDSS J142250.72+514516.49 & 11654 & -15.9 & 19100 & 7.41 & 4.0 & 2.52 & 8.46$\times10^{39}$ & 0.09 &   6500 &  2 \\
SDSS J144158.32+291434.22 & 13741 & -16.3 & 19400 & 7.47 & 4.5 & 2.80 & 1.08$\times10^{40}$ & 0.13 &  10400 &  3 \\
SDSS J080840.85+172856.48\tablenotemark{a} & 13233 & -17.1 & 19000 & 7.48 & 4.5 & 2.64 & 1.94$\times10^{40}$ & 0.18 &  18700 &  6 \\
SDSS J123944.58+145612.80 & 21534 & -17.7 & 17000 & 7.65 & 4.5 & 2.95 & 3.69$\times10^{40}$ & 0.54 &  37400 & 12 \\
SDSS J151221.08+054911.21 & 24025 & -17.6 & 15300 & 7.68 & 5.5 & 2.64 & 2.63$\times10^{40}$ & 0.32 &  40600 & 14 \\
SDSS J120955.67+142155.91 & 23340 & -17.4 & 15000 & 7.71 & 5.5 & 2.72 & 2.26$\times10^{40}$ & 0.29 &  34900 & 12 \\
SDSS J124638.82+350115.11 & 19522 & -16.8 & 16200 & 7.75 & 5.0 & 3.06 & 1.61$\times10^{40}$ & 0.26 &  24900 &  8 \\
SDSS J150316.52+111056.93 & 23405 & -18.3 & 14800 & 7.85 & 4.5 & 2.69 & 6.31$\times10^{40}$ & 0.80 &  65800 & 22 \\
SDSS J133424.53+592057.04 & 21890 & -16.8 & 14500 & 7.86 & 5.0 & 2.86 & 1.73$\times10^{40}$ & 0.23 &  27800 &  9 \\
SDSS J172955.61+534338.80 & 24308 & -18.3 & 13900 & 8.00 & 4.0 & 3.23 & 9.46$\times10^{40}$ & 1.98 &  60800 & 20 \\
        \enddata
\tablenotetext{a}{\citet{kewley07}}
 \end{deluxetable}

\end{document}